\documentclass[twocolumn,english,aps,pra,showpacs]{revtex4}
\usepackage[T1]{fontenc}
\usepackage[latin9]{inputenc}
\usepackage{amsmath}
\usepackage{graphicx}
\usepackage{amssymb}
\usepackage{amsthm}
\usepackage{multirow}

\makeatletter

\@ifundefined{textcolor}{}
{%
 \definecolor{BLACK}{gray}{0}
 \definecolor{WHITE}{gray}{1}
 \definecolor{RED}{rgb}{1,0,0}
 \definecolor{GREEN}{rgb}{0,1,0}
 \definecolor{BLUE}{rgb}{0,0,1}
 \definecolor{CYAN}{cmyk}{1,0,0,0}
 \definecolor{MAGENTA}{cmyk}{0,1,0,0}
 \definecolor{YELLOW}{cmyk}{0,0,1,0}
 }

\usepackage{txfonts}

\makeatother

\usepackage{babel}

\begin{document}

\title{Classification of arbitrary-dimensional multipartite pure states under stochastic local operations
and classical communication using the rank of coefficient matrix}

\author{Shuhao Wang,$^{1}$ Yao Lu,$^{1}$ Ming Gao,$^{1}$ Jianlian Cui,$^{2}$ and
Junlin Li$^{1,3}$\footnote{Email address:center@mail.tsinghua.edu.cn}}

\address{$^{1}$ State Key Laboratory of Low-Dimensional Quantum Physics and Department of Physics, Tsinghua University, Beijing 100084,
China\\
 $^{2}$ Department of Mathematical Sciences, Tsinghua University,
Beijing 100084, China \\
 $^{3}$ Tsinghua National Laboratory for Information Science
and Technology, Beijing 100084, China}

\date{\today }
\begin{abstract}
We study multipartite entanglement under stochastic local operations
and classical communication (SLOCC) and propose the entanglement classification
under SLOCC for arbitrary-dimensional multipartite ($n$-qudit) pure states via
the rank of coefficient matrix, together with the permutation of qudits.
The ranks of the coefficient matrices have been proved to be entanglement monotones.
The entanglement classification
of the $2 \otimes 2 \otimes 2 \otimes 4$ system is discussed
in terms of the generalized method, and 22 different SLOCC families are found.
\end{abstract}


\maketitle

Entanglement plays a vital role in quantum information processing,
which includes quantum teleportation, quantum cryptography, quantum
computation, etc \cite{book}. Classification of different types of
multipartite entanglement has been one of the main tasks in quantum
information theory. Many studies on multipartite
entanglement classification under different restrictions, such as local
operations and classical communication (LOCC) and stochastic LOCC (SLOCC) \cite{nielsen1999,bennett2000}, have been conducted in recent years.
The difference between LOCC and SLOCC can be interpreted as follows: if two states can be made equivalent up to LOCC with
some non-zero probability, they are said to be SLOCC equivalent \cite{bennett2000}.
Suppose that two \emph{n}-qudit pure states $\left|\psi\right\rangle$ and $\left|\phi\right\rangle$ are in the \emph{n}-partite
Hilbert space ${\cal H}^{n}={\cal H}_{1}\otimes{\cal H}_{2}\otimes\cdots\otimes{\cal H}_{n}$,
where ${\cal H}_{1},{\cal H}_{2},\cdots,{\cal H}_{n}$ have the dimensions
$d_{1},d_{2}\cdots,d_{n}$, respectively.
In mathematics, if $\left|\psi\right\rangle$ and $\left|\phi\right\rangle$ are LOCC equivalent iff there exists local unitary operators
${U_{(1)}},{U_{(2)}},\cdots,{U_{(n)}}$ in $U(d_1,{\Bbb C})$, $U(d_2,{\Bbb C})$, $\cdots$, $U(d_n,{\Bbb C})$, respectively, such that \cite{bennett2000}
\begin{equation}
\label{LOCC}
\left|\psi\right\rangle ={U_{(1)}}\otimes{U_{(2)}}\otimes\cdots\otimes{U_{(n)}}\left|\phi\right\rangle. \end{equation}
If $\left|\psi\right\rangle$ and $\left|\phi\right\rangle$ are SLOCC equivalent, then they can be expressed as \cite{vidal2000}
\begin{equation}
\label{SLOCC}
\left|\psi\right\rangle ={F_{(1)}}\otimes{F_{(2)}}\otimes\cdots\otimes{F_{(n)}}\left|\phi\right\rangle, \end{equation}
where ${F_{(1)}},{F_{(2)}},\cdots,{F_{(n)}}$ are invertible local operators (ILOs)
in $GL(d_1,{\Bbb C})$, $GL(d_2,{\Bbb C})$, $\cdots$, $GL(d_n,{\Bbb C})$, respectively.
In this paper, we concentrate on the entanglement classification under SLOCC.

It has been shown that two pure
states that are equivalent under SLOCC can perform the same quantum information tasks \cite{vidal2000}. The main idea of entanglement classification is to find an invariant preserved under SLOCC,
and considerable research has been conducted on the entanglement classification of three \cite{vidal2000}, four \cite{verstraete2002,lamata2006,lamata2007,zyczkowski2007,borsten2010,viehmann2011} and $n$-qubit pure 
states \cite{chen2006,bastin2009,aulbach2011,dafali2011} under SLOCC since the beginning of this century.
Recently, Li \emph{et al.} have proposed a simpler and more efficient approach for SLOCC classification of general \emph{n}-qubit
pure states in Ref. \cite{dafali2012}. A general \emph{n}-qubit pure state can be expanded as $\left| \psi \right\rangle   = \sum\nolimits_{i = 0}^{{2^n} - 1} {{a_i}\left| i \right\rangle }$, where ${a_i}$ are the coefficients and $\left| i \right\rangle $ are the binary basis states. The coefficient matrix is constructed as follows:
\begin{equation}
M(\left| \psi  \right\rangle ) = \left( {\begin{array}{*{20}{c}}
{{a_{\underbrace {0 \cdots 0}_{[n/2]}\underbrace {0 \cdots 0}_{[(n + 1)/2]}}}}& \cdots &{{a_{\underbrace {0 \cdots 0}_{[n/2]}\underbrace {1 \cdots 1}_{[(n + 1)/2]}}}}\\
{{a_{\underbrace {0 \cdots 1}_{[n/2]}\underbrace {0 \cdots 0}_{[(n + 1)/2]}}}}& \cdots &{{a_{\underbrace {0 \cdots 1}_{[n/2]}\underbrace {1 \cdots 1}_{[(n + 1)/2]}}}}\\
 \vdots & \vdots & \vdots \\
{{a_{\underbrace {1 \cdots 1}_{[n/2]}\underbrace {0 \cdots 0}_{[(n + 1)/2]}}}}& \cdots &{{a_{\underbrace {1 \cdots 1}_{[n/2]}\underbrace {1 \cdots 1}_{[(n + 1)/2]}}}}
\end{array}} \right)
\end{equation}
where the subscripts of the coefficients are written in binary form.
For two \emph{n}-qubit pure states connected by SLOCC, Li \emph{et al.} proved that the
ranks of the coefficient matrices are equal whether or not the permutation
of qubits is fulfilled on both states. This theorem
provides a way of partitioning all the \emph{n}-qubit states into different families.

With the development of quantum information theory, the importance
of qudit is gradually recognized. Maximally entangled
qudits have been shown to violate local realism more strongly and are less affected by noise
than qubits \cite{kaszlikowski2000,jchen2001,collins2002,cheng2009,son2006,he2011}. Using entangled qudits can provide
more secure scheme against eavesdropping attacks in quantum cryptography \cite{bechmann2000,bourennane2001,cerf2002,durt2003,fpan2006},
and also offers advantages
including greater channel capacity for quantum communication \cite{fujiwara2003}
as well as more reliable quantum processing \cite{ralph2007}.
Much effort has been put on the classification of bipartite and tripartite
states with higher dimensions in systems such as $2\otimes 2 \otimes n$ \cite{miyake2003,miyake2004},  $2\otimes n \otimes n$ \cite{cheng2010}, $2\otimes m \otimes n$ \cite{cornelio2006,linchen2006,eric2010} and  $m\otimes n \otimes n$ \cite{junlili2012}.

In this paper, we generalize the concept of coefficient matrix to
\emph{n}-qudit pure states. A theorem is provided to show that the rank of the coefficient matrix is invariant under SLOCC.
By calculating the rank of coefficient matrix
along with the permutation of qudits, we successfully obtain the results
of classification for \emph{n}-qudit pure states under SLOCC.
We have also proved that each of the ranks of the coefficient matrices is an entanglement monotone.
We investigate several examples and interesting entanglement properties are discovered.
Using our theorems, we discuss the entanglement classification
of the $2 \otimes 2 \otimes 2 \otimes 4$ system, which we believe has never been studied before.

Suppose an \emph{n}-qudit pure state $\left|\psi\right\rangle$ in the \emph{n}-partite
Hilbert space ${\cal H}^{n}={\cal H}_{1}\otimes{\cal H}_{2}\otimes\cdots\otimes{\cal H}_{n}$,
where ${\cal H}_{1},{\cal H}_{2},\cdots,{\cal H}_{n}$ have the dimensions
$d_{1},d_{2}\cdots,d_{n}$, respectively, which can be expanded in the form
\begin{equation}
\left|\psi\right\rangle =\sum\nolimits _{i=0}^{\prod\nolimits _{k=1}^{n}{d_{k}}-1}{{a_{i}}\left|{s_{1}}{s_{2}}\cdots{s_{n}}\right\rangle }, \end{equation}
where ${a_{i}}$ are the coefficients and $\left|{s_{1}}{s_{2}}\cdots{s_{n}}\right\rangle $
are the basis states \begin{equation}
\left|{{s_{1}}{s_{2}}\cdots{s_{n}}}\right\rangle =\left|{s_{1}}\right\rangle \otimes\left|{s_{2}}\right\rangle \otimes\cdots\otimes\left|{s_{n}}\right\rangle \end{equation}
with ${s_{k}}\in\{0,1,\cdots,{d_{k}-1}\},k=1,\cdots,n$.
The coefficient matrix $M(\left|\psi\right\rangle )$ is constructed
by arranging ${a_{i}} (i=0,\cdots,\prod\nolimits _{k=1}^{n}{d_{k}}-1)$
in lexicographical ascending order
\begin{widetext}
\begin{equation}
M(\left|\psi\right\rangle )={\left({\begin{array}{ccc}
{a_{\underbrace{0\cdots0}_{l}\underbrace{0\cdots0}_{n-l}}} & \cdots & {a_{\underbrace{0\cdots0}_{l}\underbrace{{d_{n-l}}-1\cdots{d_{n}}-1}_{n-l}}}\\
{a_{\underbrace{0\cdots1}_{l}\underbrace{0\cdots0}_{n-l}}} & \cdots & {a_{\underbrace{0\cdots1}_{l}\underbrace{{d_{n-l}}-1\cdots{d_{n}}-1}_{n-l}}}\\
\vdots & \vdots & \vdots\\
{a_{\underbrace{{d_{1}}-1\cdots{d_{l}}-1}_{l}\underbrace{0\cdots0}_{n-l}}} & \cdots & {a_{\underbrace{{d_{1}}-1\cdots{d_{l}}-1}_{l}\underbrace{{d_{n-l}}-1\cdots{d_{n}}-1}_{n-l}}}\end{array}}\right)}
\end{equation}
\end{widetext}
where $1 \le l \le n-1$.

To illustrate, we consider the \emph{n}-qudit GHZ state \cite{gllong2002}
\begin{equation}
\left|{GHZ}\right\rangle =\frac{1}{{\sqrt{d}}}({\left|0\right\rangle ^{\otimes n}}+{\left|1\right\rangle ^{\otimes n}}+\cdots+{\left|{d-1}\right\rangle ^{\otimes n}}). \end{equation}
It can be calculated that all the coefficient matrices have the form of
\begin{equation}
M(\left|{GHZ}\right\rangle)=\left( {\begin{array}{*{20}{c}}
{\frac{1}{{\sqrt d }}}&0& \cdots &0&0\\
0& \ddots & \cdots &0&0\\
 \vdots & \vdots &{\frac{1}{{\sqrt d }}}& \vdots & \vdots \\
0&0& \cdots & \ddots &0\\
0&0& \cdots &0&{\frac{1}{{\sqrt d }}}
\end{array}} \right),
\end{equation}
where the coefficient matrices are usually not square matrices, they have $d$ diagonal element being non-zero, and the non-diagonal elements are all zero.
A simple calculation shows
that $rank(\left|{GHZ}\right\rangle )=d$.

Each permutation of qubits gives a permutation $\{q_1,q_2, \cdots ,q_n\}$ of $\{1,2, \cdots ,n\}$. So in this case, the coefficient matrices $M_{q_1 \cdots q_l}$ (here we have omitted the column qudits) can be constructed by taking the corresponding permutation. The relation between all the reduced density matrices and the coefficient matrices is given by \cite{dafali2012b}
\begin{equation}
\rho _{q_{1}\cdots q_{l}}=
M_{q_{1}\cdots q_{l}} M_{q_{1}\cdots q_{l}}^{\dagger },
\end{equation}
where $M_{q_{1}\cdots q_{l}}^{\dagger }$ is the conjugate transpose of $M_{q_{1}\cdots q_{l}}$.
It is obvious that $rank(M_{q_{1}\cdots q_{l}}) = rank(\rho_{q_{1}\cdots q_{l}})$. Therefore, when considering all the particles, the local ranks \cite{vidal2000} are exactly the ranks of the coefficient matrices in the case where $l=1$.

In the following context, in the case where $l\ge2$, the permutations of qudits are included in the set
\begin{equation}
\{\sigma\}=\{(r_{1},c_{1})(r_{2},c_{2}) \cdots (r_{k},c_{k})\}\label{permutation}\end{equation}
where $1 \le {r_1} < {r_2} <  \cdots  < {r_k} < l+(n mod 2)$, $l < {c_1} < {c_2} <  \cdots  < {c_k} \le n$, and
$({r_{i}},{c_{i}})$
represents the transposition of ${r_{i}}$ and ${c_{i}}$. The purpose of choosing the permutation form in Eq. (\ref{permutation}) is to omit the permutations that end up exchanging
rows or columns in the coefficient matrix.
Letting $k$ vary from $0$ to $l-(n mod 2)$, and we get all the elements included in the set $\{\sigma\}$. The case where $k=0$ is defined as identical permutation, denoted by $\sigma_0=I$.
When $l=1$, we choose $\sigma_k=(1,k+1),k=0,1,\cdots,n-1$.

\emph{Theorem 1}.
According to Eq. (\ref{SLOCC}), the coefficient matrices of $\left|\psi\right\rangle $ and $\left|\phi\right\rangle $ satisfy the relation
\begin{eqnarray}
&&M(\left|\psi\right\rangle )= \nonumber\\
&&({F_{(1)}}\otimes\cdots\otimes{F_{([n/2])}}) M(\left|\phi\right\rangle ){({F_{([n/2]+1)}}\otimes\cdots\otimes{F_{(n)}})^{T}}.\nonumber\\
&&
\label{qudit}
\end{eqnarray}

Applying permutation $\sigma$ to both sides of Eq.
(\ref{qudit}) gives
\begin{eqnarray}
&&{M^{\sigma}}(\left|\psi\right\rangle )= \nonumber\\
&&(F^{\sigma}_{(1)}\otimes\cdots\otimes F^{\sigma}_{([n/2])}) {M^{\sigma}}(\left|\phi\right\rangle ){(F^{\sigma}_{([n/2]+1)}\otimes\cdots\otimes F^{\sigma}_{(n)})^{T}},\nonumber\\
&&
\end{eqnarray}
which indicates that ${M^{\sigma}}(\left|\psi\right\rangle )$
and ${M^{\sigma}}(\left|\phi\right\rangle )$ have the same rank.
The detailed proof is given in appendix.

Therefore, the classification of entanglement via the rank of the coefficient
matrix has the significant advantage of being independent of the dimension of
state and permutation of qudits.
Let ${{\cal F}_{n,r}}$ represents the family of all \emph{n}-qudit
states with rank \emph{r}. It is clear that all full separable states
belong to ${{\cal F}_{n,1}}$.
With the help of permutation of qudits, the families ${\cal F}_{n,r}$
can be further divided into subfamilies. Define ${\cal F}_{r}^{\sigma}$
(here we have omitted the subscript \emph{n}) as the subfamily whose
coefficient matrix rank is \emph{r} with respect to permutation
$\sigma$. The general expression of the subfamilies is
\begin{equation}
{\cal F}_{{r_1},{r_2}, \cdots {r_m}}^{{\sigma _1},{\sigma _2}, \cdots ,{\sigma _m}} = {\cal F}_{{r_1}}^{{\sigma _1}} \cap  \cdots  \cap {\cal F}_{{r_m}}^{{\sigma _m}}.
\end{equation}

In order to
maximize the number of families, the value of $l$ is given by \begin{equation}
l=\rm{argmax}\{{\cal P}(\emph{l})\},\end{equation}
where\begin{equation}
{\cal P}(l) = \prod\limits_{\{ \sigma \} } {\min \{ \prod\nolimits_{k = 1}^l {{d_{q_k}},} \prod\nolimits_{k = l + 1}^n {{d_{q_k}}\} } } \end{equation}
with $d_{q_k}$ the dimension of the party corresponding to $q_k$.
It is obvious that for states with each party of the same dimension, the family number is maximized when $l=[n/2]$.

\emph{Theorem 2.}
Each of the ranks of the coefficient matrices is an entanglement monotone.

\emph{Proof.}
It has been shown that the rank of the coefficient matrice $M_{{q_1},{q_2}, \cdots ,{q_l}}(\left| {{\psi}} \right\rangle)$, which is the direct generalization of the Schmidt rank of the bipartite pure states, cannot be increased by LOCC \cite{lo2001}.
Therefore, $rank(M_{{q_1},{q_2}, \cdots ,{q_l}}(\left| {{\psi}} \right\rangle))$ is an entanglement monotone.

The theorem has shown that the
rank of coefficient matrix is closely connected with the degree of entanglement.

As an application of the generalized method, consider the following state:
\begin{eqnarray}
&&\left|{{l_{1}},{l_{2}},n}\right\rangle = \nonumber\\
&&{\left({\frac{{n!}}{{{l_{0}}!{l_{1}}!{l_{2}}!}}}\right)^{-\frac{1}{2}}}\sum\limits _{k}{{P_{k}}\left|{\underbrace{1,\cdots,1}_{{l_{1}}},\underbrace{2,\cdots,2}_{{l_{2}}},\underbrace{0,\cdots,0}_{{l_{0}}}}\right\rangle },
\label{d3}
\end{eqnarray}
where $\left|1\right\rangle ,\left|2\right\rangle $ are the excitations,
$\left|0\right\rangle $ represents the ground state, and ${l_{0}},{l_{1}},{l_{2}}$
are the number of states $\left|0\right\rangle ,\left|1\right\rangle ,\left|2\right\rangle $,
respectively, which satisfy ${l_{1}}+{l_{2}}\le n-1$. $\{{P_{k}}\}$
is the set that contains all permutations.
We denote the states in Eq. (\ref{d3}) as $D_3^n$ states.

For $D_3^n$ states,
states $\left|{{l_{1}},{l_{2}},n}\right\rangle$,
$\left|{{l_{2}},{l_{1}},n}\right\rangle$,
$\left|{{n-l_{1}-l_{2}},{l_{1}},n}\right\rangle$,
$\left|{{n-l_{1}-l_{2}},{l_{2}},n}\right\rangle$,
$\left|{{l_{1}},{n-l_{1}-l_{2}},n}\right\rangle$, and
$\left|{{l_{2}},{n-l_{1}-l_{2}},n}\right\rangle$ can be transformed into each other under SLOCC,
namely, they belong to the same family. In the following, we can arrange these states and denote them as $a(l_1,l_2,l_0)$, where $l_0=n-l_1-l_2$.
We study the classification of entanglement of $D_3^9$ states
with respect to ${l_{1}}$, ${l_{2}}$ and $l_{0}$.
The variance of ${l_{1}}$, ${l_{2}}$ and $l_{0}$
and the ranks of the coefficient matrices $M_{q_1q_2q_3q_4}$ under different arrangements are shown in Fig. 1, which shows that the rank of the coefficient matrix increases with the decrease of the variance, and most of the $D_3^9$ states can be distinguished by the ranks of the coefficient matrices.

Physically speaking, states $\left|0\right\rangle $, $\left|1\right\rangle$ and $\left|2\right\rangle $ are on an equal footing. So the state is maximal entangled when $l_0$, $l_1$ and $l_2$ are close to each other, namely, the variance of $l_0$, $l_1$ and $l_2$ is as small as possible. According to Theorem 2, Fig. 1 shows an inverse relationship between the variance and the rank of $M_{q_1q_2q_3q_4}$.

\begin{figure}[!h]

\begin{centering}
\includegraphics[width=8.5cm]{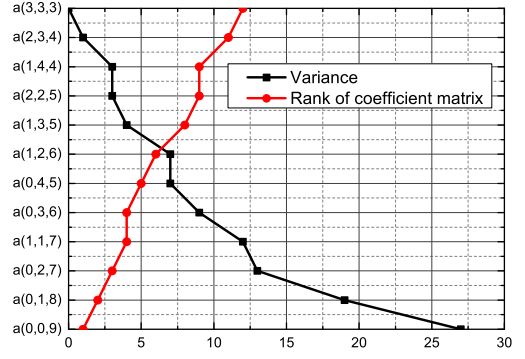}
\caption{(Color online) Variance of ${l_{1}}$, ${l_{2}}$ and $l_{0}$
and ranks of the coefficient matrices under different arrangements (shown in the vertical axis) existing in $D_3^9$ states.}

\par\end{centering}

\centering{}\label{fig1}
\end{figure}

We then consider $D_4^n$ states. which are defined as
\begin{eqnarray}
&&\left|{{l_{1}},{l_{2}},{l_{3}},n}\right\rangle = \nonumber\\
&&{\left({\frac{{n!}}{{{l_{0}}!{l_{1}}!{l_{2}}!{l_{3}}!}}}\right)^{-\frac{1}{2}}}\sum\limits _{k}{{P_{k}}\left|{\underbrace{1,\cdots,1}_{{l_{1}}},\underbrace{2,\cdots,2}_{{l_{2}}},\underbrace{3,\cdots,3}_{{l_{3}}},\underbrace{0,\cdots,0}_{{l_{0}}}}\right\rangle },\nonumber\\
&&
\end{eqnarray}
where $\left|1\right\rangle ,\left|2\right\rangle $ and $\left|3\right\rangle $
are the excitations with ${l_{1}},{l_{2}}$ and ${l_{3}}$ as their
numbers, which satisfy ${l_{1}}+{l_{2}}+{l_{3}}\le n-1$, whereas $\left|0\right\rangle $
is the ground state.

We study the classification of entanglement of $D_4^8$ states
with respect to ${l_{1}}$, ${l_{2}}$, ${l_{3}}$, and $l_{0}$.
The variance of ${l_{1}}$, ${l_{2}}$, ${l_{3}}$ and $l_{0}$
and the ranks of the coefficient matrices $M_{q_1q_2q_3q_4}$ under different
arrangements are shown in Fig. 2.
The rank of the coefficient matrices shows a contrasting trend with the decrease of the variance, the physical interpretation of this phenomenon is the same as the $D_3^n$ states,
and we can distinguish most states in terms of the ranks of the coefficient matrices.

\begin{figure}[!h]

\begin{centering}
\includegraphics[width=8.5cm]{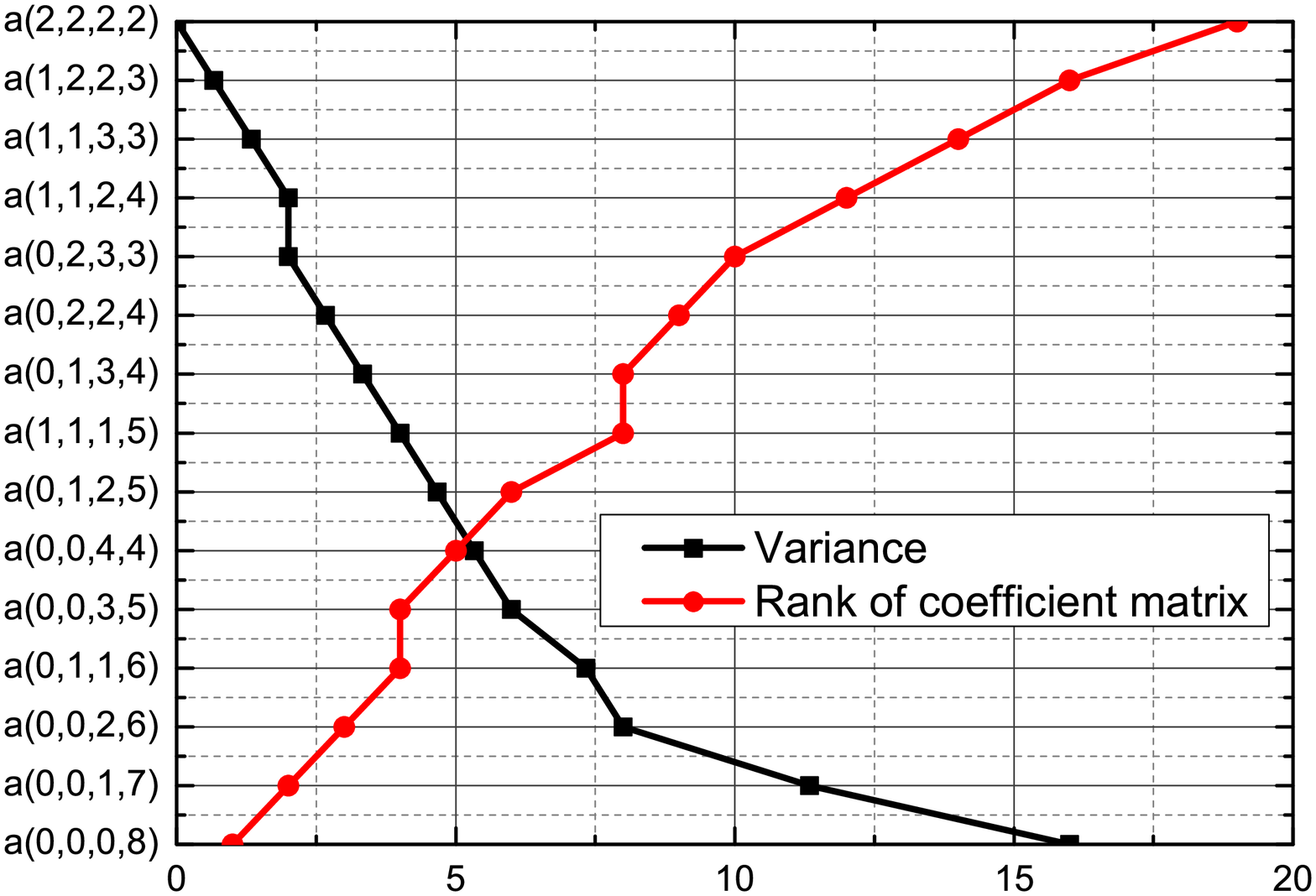}
\caption{(Color online) Variance of ${l_{1}}$, ${l_{2}}$, ${l_{3}}$ and $l_{0}$
and ranks of coefficient matrices $M_{q_1q_2q_3q_4}$ under different arrangements (shown in the vertical axis)
existing in $D_4^8$ states.}

\par\end{centering}

\centering{}\label{fig2}
\end{figure}

In the end, we discuss the entanglement classification of the $2 \otimes 2 \otimes 2 \otimes 4$ system.
For the cases where $l=1$, $l=2$, and $l=3$, the values of ${\cal P}(l)$
are $4$, $64$ and $4$, respectively. To maximize the family number, we consider the case where $l=2$.
The set of permutation consists of three elements:
$\{\sigma\}=\{\sigma_{0}=I,\sigma_{1}=(1,3),\sigma_{2}=(1,4)\}$.
The classification results are shown in Table I.
It needs to be noted that the entangled states ($\left|W\right\rangle$ and $\left|GHZ\right\rangle$ states) in ${\cal F}_{2,2,2}^{\sigma_{0},\sigma_{1},\sigma_{2}}$
have a similar Frobenius algebra structure \cite{coecke2010}. The entanglement structure of the $2 \otimes 2 \otimes 2 \otimes 4$ system is illustrated by an entanglement pyramid in Fig. 3.

\begin{table}[!h]
\tabcolsep 0pt
\caption{SLOCC classification of the $2 \otimes 2 \otimes 2 \otimes 4$ system. The permutations are $\sigma_{0}=I,\sigma_{1}=(1,3),\sigma_{2}=(1,4)$.}
\vspace*{-12pt}
\begin{center}
\def\temptablewidth{0.5\textwidth}
{\rule{\temptablewidth}{1pt}}
\begin{tabular*}{\temptablewidth}{@{\extracolsep{\fill}}cc}

SLOCC family & Representative entangled states \\
\hline
${\cal F}_{4,4,4}^{\sigma_{0},\sigma_{1},\sigma_{2}}$ & $\left|0000\right\rangle +\left|0010\right\rangle + \left|0101\right\rangle + \left|0111\right\rangle$
 \\ & $+ \left|1002\right\rangle+ \left|1012\right\rangle+ \left|1103\right\rangle+ \left|1113\right\rangle  $ \\
\hline
${\cal F}_{4,4,3}^{\sigma_{0},\sigma_{1},\sigma_{2}}$ & $\left|0000\right\rangle +\left|1010\right\rangle + \left|1001\right\rangle+ \left|0102\right\rangle + \left|1113\right\rangle  $\\
\hline
${\cal F}_{4,3,4}^{\sigma_{0},\sigma_{1},\sigma_{2}}$ & $\left|0000\right\rangle +\left|0110\right\rangle +\left|1100\right\rangle +\left|1002\right\rangle + \left|1113\right\rangle  $\\
\hline
${\cal F}_{3,4,4}^{\sigma_{0},\sigma_{1},\sigma_{2}}$ & $\left|0000\right\rangle +\left|0110\right\rangle + \left|1100\right\rangle+ \left|0012\right\rangle + \left|1113\right\rangle  $ \\
\hline
${\cal F}_{4,3,3}^{\sigma_{0},\sigma_{1},\sigma_{2}}$ & $\left|0000\right\rangle +\left|0111\right\rangle + \left|1012\right\rangle + \left|1113\right\rangle  $\\
\hline
${\cal F}_{3,4,3}^{\sigma_{0},\sigma_{1},\sigma_{2}}$ & $\left|0000\right\rangle +\left|1101\right\rangle + \left|1012\right\rangle + \left|1113\right\rangle  $\\
\hline
${\cal F}_{3,3,4}^{\sigma_{0},\sigma_{1},\sigma_{2}}$ & $\left|0000\right\rangle +\left|0111\right\rangle + \left|1102\right\rangle + \left|1113\right\rangle  $\\
\hline
${\cal F}_{4,4,2}^{\sigma_{0},\sigma_{1},\sigma_{2}}$ & $\left|0000\right\rangle +\left|1010\right\rangle + \left|0102\right\rangle + \left|1113\right\rangle  $\\
\hline
${\cal F}_{4,2,4}^{\sigma_{0},\sigma_{1},\sigma_{2}}$ & $\left|0000\right\rangle +\left|0110\right\rangle + \left|1002\right\rangle + \left|1113\right\rangle  $\\
\hline
${\cal F}_{2,4,4}^{\sigma_{0},\sigma_{1},\sigma_{2}}$ & $\left|0000\right\rangle +\left|1100\right\rangle + \left|0012\right\rangle + \left|1113\right\rangle  $\\
\hline
${\cal F}_{3,3,3}^{\sigma_{0},\sigma_{1},\sigma_{2}}$ & $\left|0000\right\rangle +\left|1010\right\rangle +\left|1001\right\rangle+ \left|1113\right\rangle  $\\
\hline
${\cal F}_{3,3,2}^{\sigma_{0},\sigma_{1},\sigma_{2}}$ & $\left|0000\right\rangle +\left|1010\right\rangle + \left|1112\right\rangle $\\
\hline
${\cal F}_{3,2,3}^{\sigma_{0},\sigma_{1},\sigma_{2}}$ & $\left|0000\right\rangle +\left|1001\right\rangle + \left|1112\right\rangle $\\
\hline
${\cal F}_{2,3,3}^{\sigma_{0},\sigma_{1},\sigma_{2}}$ & $\left|0000\right\rangle +\left|1100\right\rangle + \left|1112\right\rangle  $\\
\hline
 & $\left|1010\right\rangle +\left|1100\right\rangle+\left|1001\right\rangle  $\\

${\cal F}_{2,2,2}^{\sigma_{0},\sigma_{1},\sigma_{2}}$ & $\left|0001\right\rangle +\left|0010\right\rangle +\left|0100\right\rangle +\left|1000\right\rangle $\\

 & $\left|0000\right\rangle +\left|1111\right\rangle $\\
\hline
${\cal F}_{4,4,1}^{\sigma_{0},\sigma_{1},\sigma_{2}}$ & $\left|0000\right\rangle +\left|0011\right\rangle+\left|1100\right\rangle+\left|1111\right\rangle $\\
\hline
${\cal F}_{4,1,4}^{\sigma_{0},\sigma_{1},\sigma_{2}}$ & $\left|0000\right\rangle +\left|1001\right\rangle+\left|0110\right\rangle+\left|1111\right\rangle $\\
\hline
${\cal F}_{1,4,4}^{\sigma_{0},\sigma_{1},\sigma_{2}}$ & $\left|0000\right\rangle +\left|1010\right\rangle+\left|0101\right\rangle+\left|1111\right\rangle $\\
\hline
${\cal F}_{2,2,1}^{\sigma_{0},\sigma_{1},\sigma_{2}}$ & $\left|1100\right\rangle +\left|1001\right\rangle $\\
\hline
${\cal F}_{2,1,2}^{\sigma_{0},\sigma_{1},\sigma_{2}}$ & $\left|1100\right\rangle +\left|1010\right\rangle $\\
\hline
${\cal F}_{1,2,2}^{\sigma_{0},\sigma_{1},\sigma_{2}}$ & $\left|1010\right\rangle +\left|1001\right\rangle $\\
\hline
${\cal F}_{1,1,1}^{\sigma_{0},\sigma_{1},\sigma_{2}}$ & $\left|0000\right\rangle   $\\

       \end{tabular*}
       {\rule{\temptablewidth}{1pt}}
       \end{center}
       \end{table}

\begin{figure}[!h]

\begin{centering}
\includegraphics[width=8.5cm]{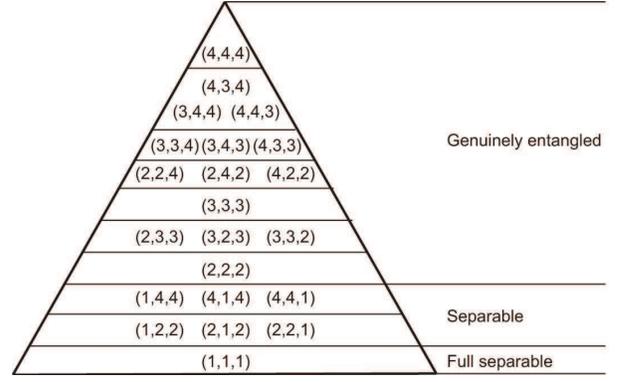}
\caption{The entanglement pyramid of the $2 \otimes 2 \otimes 2 \otimes 4$ system, where we use $(i,j,k)$ to represent ${\cal F}_{i,j,k}^{\sigma_{0},\sigma_{1},\sigma_{2}}$.}

\par\end{centering}

\centering{}\label{fig3}
\end{figure}

In summary, the rank invariance of the coefficient matrix under SLOCC has been proven to be
valid in the \emph{n}-qudit pure states regardless of the dimension of
each partite and the permutation of qudits.
It has also been proved that each of the ranks of the coefficient matrices is an entanglement monotone.
Numerical results showed
that this generalization can investigate the entanglement
feature of quantum states with qudits.
We have discussed the entanglement classification of the $2 \otimes 2 \otimes 2 \otimes 4$ system
and found 22 different SLOCC families with respect to the generalized method.
We expect that our generalization could come up with further
theoretical and experimental results.

This work was supported by the National Natural Science Foundation
of China (Grant Nos. 11175094 and 11271217) and the National Basic Research Program
of China (2009CB929402, 2011CB9216002).

\section*{APPENDIX}

Now we prove the following theorem:

Let $|\psi\rangle $, $|\phi\rangle $ be any states in the $n$-partite Hilbert space ${\mathcal H}={\mathcal H}_1\otimes {\mathcal H}_2\otimes \cdots \otimes {\mathcal H}_n$, where ${\mathcal H}_i$ is of dimension $d_i$, $1\leq i\leq n$. If there exist $A_i\in {\mathcal M}_{d_i}({\mathbb C})$ ($1\leq i\leq n$) such that
\begin{equation}
|\psi \rangle =A_1\otimes A_2\otimes \cdots \otimes A_n|\phi \rangle ,
\end{equation}
then, for any $1\leq l<n$,
\begin{equation}
M(|\psi \rangle )=A_1\otimes \cdots \otimes A_lM(|\phi \rangle )(A_{l+1}\otimes \cdots \otimes A_n)^T.
\label{genetheorem}
\end{equation}

We will prove Eq. (\ref{genetheorem}) by the induction method. Clearly, if $A_i=I_i$ (the identity matrix in ${\mathcal M}_{d_i}({\mathbb C})$) for every $1\leq i\leq n$, then equation Eq. (\ref{qudit}) holds.

Let $|\psi\rangle =\sum _{i=0}^{d_1\cdots d_n -1}c_i|i\rangle $ and for $1\leq r<n$,
\begin{equation}
|\psi \rangle =I_1\otimes \cdots \otimes I_r\otimes A_{r+1}\otimes \cdots \otimes A_n|\phi \rangle .
\end{equation}

For any $1\leq l<n$, we assume that
\begin{eqnarray}
M(|\psi \rangle )&=&I_1\otimes \cdots \otimes I_r\otimes A_{r+1}\nonumber\\
&&\otimes \cdots \otimes A_l M(|\phi \rangle )(A_{l+1}\otimes \cdots \otimes A_n)^T,\nonumber\\
\rm{when} && r+1\leq l<n;\nonumber\\
M(|\psi \rangle )&=&I_1\otimes \cdots \otimes I_l M(|\phi \rangle )\nonumber\\
&& \times (I_{l+1}\otimes \cdots \otimes I_r\otimes A_{r+1}\otimes \cdots \otimes A_n)^T,\nonumber\\
\rm{when} && 1\leq l<r< n;\nonumber\\
M(|\psi \rangle )&=&I_1\otimes \cdots \otimes I_l M(|\phi \rangle )(A_{r+1}\otimes \cdots \otimes A_n)^T,\nonumber\\
\rm{when} && 1\leq l=r< n.
\end{eqnarray}

Next, we will prove that when
\begin{equation}
|\psi ^\prime\rangle = I_1\otimes \cdots \otimes I_{r-1}\otimes A_{r}\otimes \cdots \otimes A_n|\phi \rangle ,
\end{equation}
there is
\begin{eqnarray}
M(|\psi ^\prime \rangle )&=&I_1\otimes \cdots \otimes I_{r-1}\otimes A_r\otimes \cdots \otimes A_l \nonumber\\
&&\times M(|\phi \rangle )(A_{l+1}\otimes \cdots \otimes A_n)^T,\nonumber\\
\rm{when} &&r+1\leq l<n;\nonumber \\
M(|\psi ^\prime \rangle )&=&I_1\otimes \cdots \otimes I_l M(|\phi \rangle )\nonumber\\
&&\times (I_{l+1}\otimes \cdots \otimes I_{r-1}\otimes A_r\otimes \cdots \otimes A_n)^T,\nonumber\\
\rm{when} &&1\leq l<r<n;\nonumber \\
M(|\psi ^\prime \rangle )&=&I_1\otimes \cdots \otimes I_{r-1}\otimes A_r M(|\phi \rangle )(A_{r+1}\otimes \cdots \otimes A_n)^T,\nonumber\\
\rm{when} &&1\leq l=r<n.\nonumber \\
\end{eqnarray}

Write $|\psi ^\prime \rangle =\sum _{i=0}^{d_1\cdots d_n -1}b_i|i\rangle $ and
\begin{equation}
A_r=\left (\begin{array}{cccc}
a_{11} & a_{12} & \cdots & a_{1d_r}\\
a_{21} & a_{22} & \cdots & a_{2d_r}\\
\vdots & \vdots & \cdots & \vdots \\
a_{d_r1} & a_{d_r2} & \cdots & a_{d_rd_r}\end{array}\right ).
\end{equation}

Since
\begin{equation}
|\psi ^\prime \rangle =I_1\otimes \cdots \otimes I_{r-1}\otimes A_r\otimes I_{r+1} \otimes \cdots \otimes I_n|\psi \rangle ,
\label{since}
\end{equation}
we need only prove that
\begin{eqnarray}
M(|\psi ^\prime \rangle ) & = & I_1\otimes \cdots \otimes I_{r-1}\otimes A_r\otimes I_{r+1} \otimes \cdots \otimes I_l M(|\psi \rangle ),\nonumber\\
\rm{when} & &r+1\leq l<n;\nonumber\\
M(|\psi ^\prime \rangle )& = &I_1\otimes \cdots \otimes I_l M(|\psi \rangle )\nonumber\\
 & &\times (I_{l+1}\otimes \cdots \otimes I_{r-1}\otimes A_r\otimes I_{r+1}\otimes \cdots \otimes I_n)^T,\nonumber\\
\rm{when} & &1\leq l<r;\nonumber\\
M(|\psi ^\prime \rangle )& = &I_1\otimes \cdots \otimes I_{r-1}\otimes A_r M(|\psi \rangle ),\nonumber\\
\rm{when} & &1\leq r=l<n.
\label{first}
\end{eqnarray}

From Eq. (\ref{since}), it can be computed that
\begin{eqnarray}
b_{khd_r+s+(t-1)h}&=&a_{t1}c_{khd_r+s}+a_{t2}c_{khd_r+h+s}\nonumber\\
&&+\cdots +a_{td_r}c_{khd_r+(d_r-1)h+s},
\label{second}
\end{eqnarray}
where $t=1, 2, \ldots , d_r$, $k=0, 1, \ldots , d_1\cdots d_{r-1}-1$, $s=0, 1, \ldots , d-1$, $h=d_{r+1}\cdots d_{n}$. If $r+1\leq l<n$, write
\begin{equation}
M(|\psi \rangle )=\left (\begin{array}{cccccc}
c_0 & c_1 & \cdots & c_{d_{l+1}\cdots d_n-1}\\
c_{d_{l+1}\cdots d_n} & c_{d_{l+1}\cdots d_n+1} & \cdots & c_{2d_{l+1}\cdots d_n-1}\\
\vdots & \vdots & \vdots & \vdots \\
c_d & c_{h+1} & \cdots & c_{h+d_{l+1}\cdots d_n-1}\\
\vdots & \vdots & \vdots & \vdots \\
c_{(d_1\cdots d_l-1)d_{l+1}\cdots d_n } & c_{(d_1\cdots d_l-1)d_{l+1}\cdots d_n +1} & \cdots & c_{d_1\cdots d_n-1}\end{array}\right );
\end{equation}
if $1\leq l<r<n$, write
\begin{widetext}
\begin{equation}
M(|\psi \rangle )=\left (\begin{array}{cccccc}
c_0 &  c_1 & \cdots & c_h & \cdots & c_{d_{l+1}\cdots d_n-1}\\
c_{d_{l+1}\cdots d_n} & c_{d_{l+1}\cdots d_n+1} & \cdots & c_{d_{l+1}\cdots d_n+d_{r+1}\cdots d_n} &  \cdots & c_{2d_{l+1}\cdots d_n-1}\\
\vdots & \vdots & \vdots & \vdots & \vdots & \vdots \\
c_{(d_1\cdots d_l-1)d_{l+1}\cdots d_n } & c_{(d_1\cdots d_l-1)d_{l+1}\cdots d_n +1} & \cdots & c_{(d_1\cdots d_l-1)d_{l+1}\cdots d_n +h} & \cdots & c_{d_1\cdots d_n-1}\end{array}\right );
\end{equation}
\end{widetext}
if $1\leq l=r<n$, write
\begin{equation}
M(|\psi \rangle )=\left (\begin{array}{cccc}
c_0 &  c_1 & \cdots & c_{h-1}\\
c_{d} & c_{d+1} & \cdots & c_{2h-1}\\
\vdots & \vdots & \vdots & \vdots \\
c_{(d_1\cdots d_l-1)h} & c_{(d_1\cdots d_l-1)h+1} & \cdots & c_{d_1\cdots d_n-1}\end{array}\right ),
\end{equation}
then it follows from Eq. (\ref{second}) that equations Eq. (\ref{first}) holds.

Finally, we consider the permutation of qudits.
Applying the permutation $\sigma$ defined in Eq. (\ref{permutation}) to both sides of Eq. (\ref{genetheorem}) and we have
\begin{eqnarray}
{M^\sigma }(\left| \psi  \right\rangle ) &=& A_1^\sigma  \otimes  \cdots  \otimes A_l^\sigma  \nonumber\\
&&\times {M^\sigma }(\left| \phi  \right\rangle ){(A_{l + 1}^\sigma  \otimes  \cdots  \otimes A_n^\sigma )^T}.
\label{genepermutation}
\end{eqnarray}
When ${A_1},\cdots,{A_n}$ are ILOs,
it can be directly concluded from Eq. (\ref{genepermutation}) that ${M^\sigma }(\left| \psi  \right\rangle )$ and ${M^\sigma }(\left| \phi  \right\rangle )$ have the same rank.
Thus two SLOCC
equivalent states have the same rank with respect to every
permutation of qudits.

\end{document}